\documentclass[12pt]{IEEEtran}
\usepackage{setspace}

\usepackage{cite}
\usepackage[final]{graphicx}

\usepackage{amssymb}
\usepackage{amsmath}
\usepackage{enumerate}
\usepackage{epstopdf}
\usepackage{subfigure}
\usepackage{lipsum}
\usepackage{xcolor}
\usepackage{tikz}
\usepackage{lipsum}
\usepackage[]{caption}
\begin{document}

\title{Multi-Gigabit Wireline Systems over Copper: An Interference Cancellation Perspective}
\author{ S.~M.~ Zafaruddin,~\IEEEmembership{Member,~IEEE} and
	Amir Leshem,~\IEEEmembership{Senior Member,~IEEE} 
	
\thanks{ S.~M.~Zafaruddin is with Department of Electrical and Electronics Engineering, BITS Pilani, Pilani-333031, India (email: syed.zafaruddin@pilani.bits-pilani.ac.in).}
\thanks{ Amir Leshem  is Faculty of Engineering, Bar-Ilan University, Ramat Gan 52900, Israel  (email: leshema@biu.ac.il).}
	
}

\maketitle

\begin{abstract}
	   Interference cancellation is the main driving technology in enhancing the transmission rates over telephone lines above $100$ Mbps.  Still,  crosstalk interference in  multi-pair digital subscriber line (DSL) systems at higher frequencies has not been dealt with sufficiently.   The upcoming G.(mg)fast DSL system envisions the use of extremely high bandwidth and full-duplex transmissions generating significantly  higher crosstalk  and self-interference signals. More powerful interference cancellation techniques are required to  enable multi-gigabit per second data rate transmission over copper lines. In this  article, we analyze the performance of interference cancellation techniques, with a focus on novel research approaches and design considerations for efficient interference mitigation for multi-gigabit transmission over standard copper lines.  We also detail  novel approaches for interference cancellation in the upcoming technologies.

\end{abstract}

	\begin{table*}
			\begin{center}
		{\includegraphics[scale=0.5]{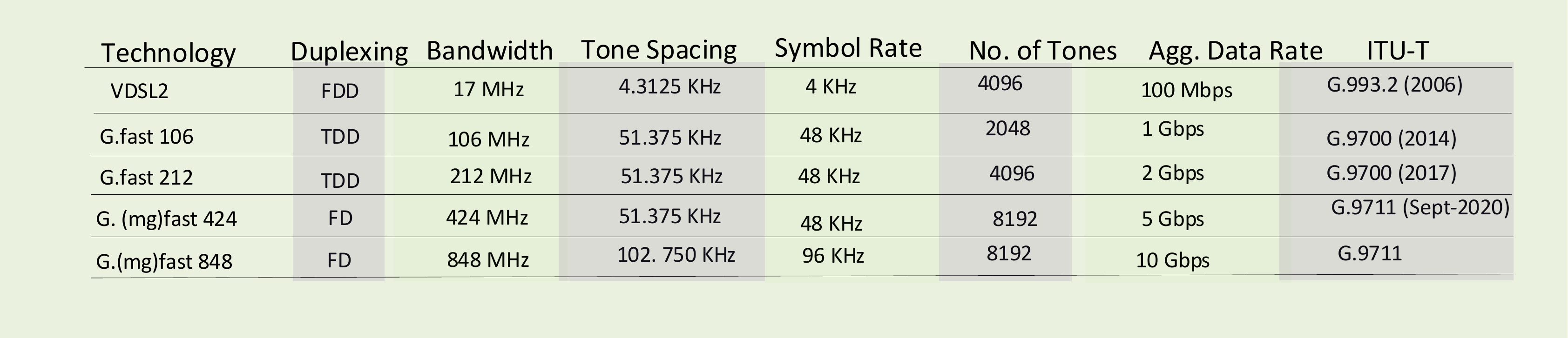}}	
			\caption{Evolution of transmission over telephone lines technology with system parameters.  Aggregate data rate is for each user upstream and downstream combined for FDD and TDD systems, and either upstream or downstream  in the FD system. FD: full-duplex; FDD: frequency division duplexing; TDD: time division duplexing.}	
	\end{center}
\label{fig:overview}
\end{table*}

\section{Introduction}
Evolution of data transmission technologies over copper lines has been vary rapid. Recent extensions achieve   bit rates of several gigabits per second \cite{g9701,timmers2013mag,Coomans2015,zafar2017spm}.  This growth is possible by exploiting higher transmission bandwidth over the same telephone lines \cite{Oksman2019}. However, an increase in the bandwidth increases the crosstalk among closely packed telephone lines in a binder caused by  the electromagnetic coupling.    Furthermore, the upcoming copper line technology might include  full-duplex transmission creating problems of self-interference \cite{Laneer2017fd,Strobel2018}. In addition, subscribers can experience interference external to the binder. Novel interference cancellation techniques are required to realize a multi-gigabit rate transmission over copper lines.  In this  article, we analyze the performance of interference cancellation techniques, with a focus on novel research approaches  for efficient interference mitigation for next generation digital subscriber line (DSL) systems.

The coordinated processing of the signals over all lines referred to as vectoring \cite{ginis2002}, is a viable technique to deal with crosstalk in the DSL systems. This joint processing can be applied on the signals transmitted from the central point to the end users  in the downstream and on the signal received at the central point from the users in the upstream transmissions. Since  DSL channels are well conditioned at lower frequencies,
linear vectored processing is near optimal for crosstalk cancellation in the very high-speed digital subscriber line (VDSL) system \cite{cendrillon2006near}, and it is recommended for the $106$-MHz G.fast system \cite{g9701}. With further increase in frequency as in the $212$-MHz G.fast system, and in the upcoming  $424$-MHz and $848$-MHz G.(mg)fast systems,  crosstalk is significantly higher than the direct path thereby the channel matrices are no longer diagonal dominant. In   the $212$-MHz G.fast system, linear cancellation schemes are shown  to be near-optimal \cite{Lanneer2017}. Recently,  lattice reduction based techniques are employed for the  the $212$-MHz G.fast system that incurs an extra run-time complexity \cite{Zhang2020}. It is interesting to evaluate the performance of linear and non-linear processing for G.(mg)fast systems.  It is well known that the nonlinear schemes are computationally more complex than the linear techniques. This opens an exciting opportunity to develop computationally efficient crosstalk cancellation schemes for the next generation multi-pair DSL systems. 

The upcoming G.(mg)fast systems promise to  achieve multi-gigabit data rates using full-duplex (FD) transmissions, and by exploiting higher bandwidth  than the previous standards \cite{Laneer2017fd,Strobel2018,Oksman2019}.  This is in contrast to the VDSL system based on the frequency division duplexing (FDD) and  to the G.fast system based on the time division duplexing (TDD). Although FD transmissions can double the throughput, self-interference or the echo signal can be detrimental for the data rate performance. It is interesting to note that the symmetric  DSL technologies were based on the FD where the self-interference was dealt with an echo canceler \cite{g9912}. However, in the central office (CO) topology,  near-end crosstalk (NEXT) beyond $600$ \mbox{kHz} was prohibitive, and later standards used either FDD or TDD. In this context, the performance analysis of existing echo cancelers in the G.(mg)fast systems and development of novel techniques tailored for  multi-gigabit data rate transmissions are desirable. NEXT cancelling was used and is still relevant in point to point multi-input-multi-output (MIMO) topologies such as certain multi-pair symmetric high-speed DSL  (SHDSL) or $1/10$ \mbox{Gbps} Ethernet. Unfortunately, in point to multi-point, NEXT cancellation at the CPE is hard because of the separation of the receivers. It has been suggested that for certain topologies, proper power management of the signals can overcome this problem \cite{Laneer2017fd}.

 The lack of signal-coordination of the interference external to binder i.e., from other existing broadband services and through the usage of various electrical appliances in the vicinity of a CPE limits the possibility of its cancellation.  The usual practice to deal with the impulse noise is the use of error-correcting codes together and physical layer retransmission. The use of noise reference modules has shown promise for impulse noise mitigation in VDSL system \cite{zafar_icc2016}.

This article analyzes the performance of multi-gigabit rate DSL systems.  In this paper our focus is on comparative study of FEXT cancellation (vectoring) techniques, which are the main building block in VDSL and G.fast and will continue to be significant in G.(mg)fast. 

\begin{figure*}[t]
	\begin{center}	
		\subfigure{\includegraphics[width=\columnwidth]{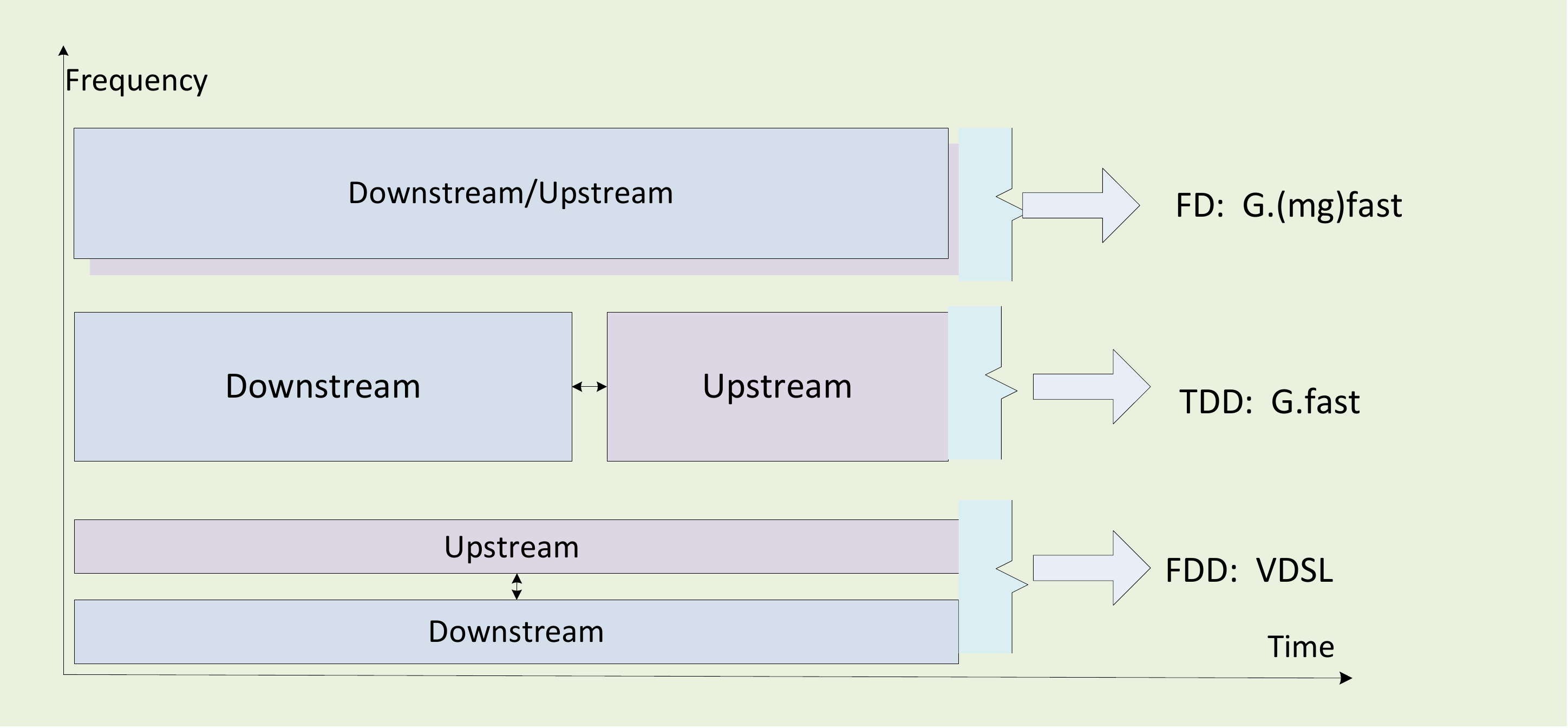}}
		\subfigure{\includegraphics[width=8.5cm,height=4.3cm]{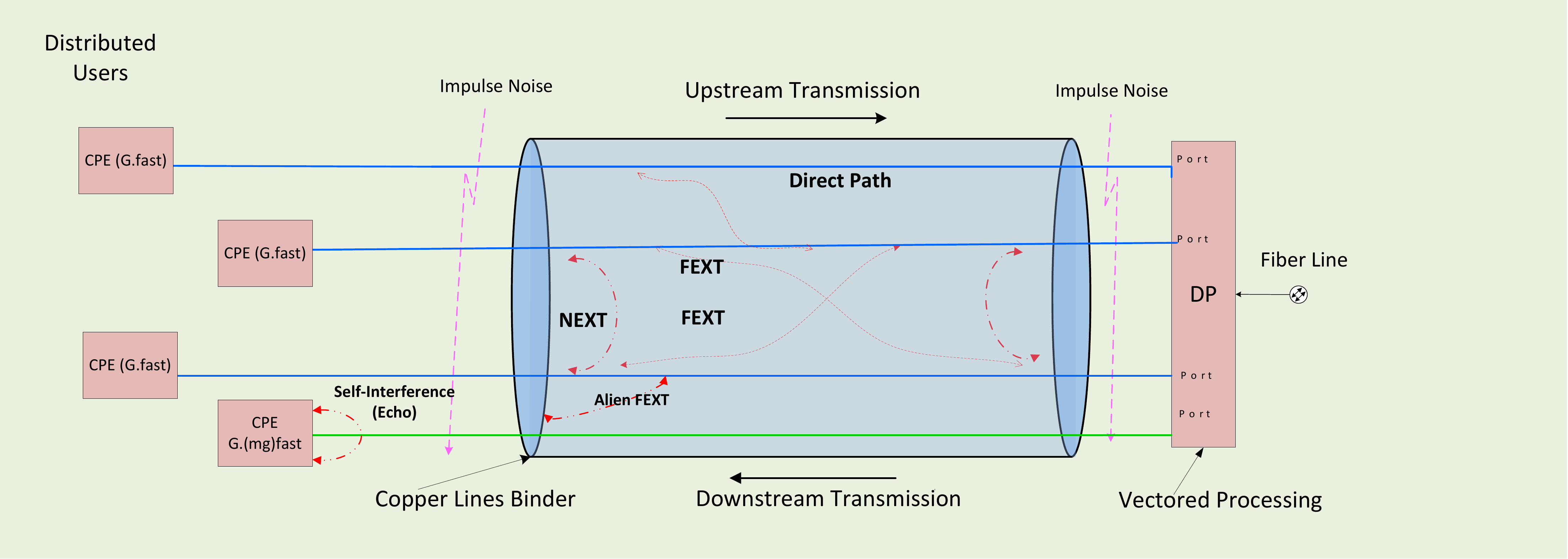}}
		\caption{Duplexing techniques [left] and interference environment [right]  in a multi-pair DSL system.  FD: full-duplex; FDD: frequency division duplexing; TDD: time division duplexing; NEXT: near-end crosstalk; FEXT: far-end crosstalk; DP: distribution point; CPE: customer premise equipment.}
		\label{fig:crosstalk_duplexing}
	\end{center}
\end{figure*}

\section{Overview of DSL Technologies}
	Discrete multi-tone (DMT) is the de facto multi-carrier modulation for frequency selective DSL channels. Of-course, there is a change in the parameters for different technologies. The main system parameter is the sub-carrier spacing which is chosen to limit the number of tones to $4096$ to accomplish per-tone processing. For the VDSL system with $17$ MHz bandwidth, the sub-carrier spacing is $4.3125$ \mbox{KHz}. The sub-carrier spacing in the G.fast is $51.75$ \mbox{KHz} which accounts for  $2048$ tones and $4096$ tones in the $106$ \mbox{MHz} and $212$ \mbox{MHz} G.fast systems, respectively. The $424$ \mbox{MHz}  G.(mg)fast system employs 8192 tones with a subcariier spacing of $51.75$ \mbox{KHz}.  In order to limit the number of tones, we anticipate that the sub-carrier spacing of  the G.(mg)fast- $848$ \mbox{MHz} system  should be double compared  the  G.fast system. 
	
	Duplexing techniques are another distinguishing  feature among different DSL technologies.  These techniques  separate the upstream and downstream to	avoid the NEXT and echo signals in the DSL systems. The VDSL standard separates the upstream and downstream	in the frequency domain, known as the FDD. The latest G.fast standard  employs a TDD where the upstream and	downstream are transmitted at different times. The next generation G.(mg)fast targets a more ambitious full duplex transmission to double the throughput with a provision of  efficient echo and NEXT cancellation.  Table 1 presents important parameters of the various technologies.

\section{Noise and Interference Environment}
Thermal noise caused by the random movement of electrons is inherently present in all communication systems. This presents a noise floor which is generally considered with a power spectral density (PSD) of  $-174$ \mbox{dBm/Hz} at the room temperature. The PSD of background noise is around $-140$ \mbox{dBm/Hz} for the VDSL frequencies which decreases to $-150$ \mbox{dBm/Hz} for the G.fast system. The noise floor for the G.(mg)fast systems needs to be measured.

Crosstalk interference occurs when several unshielded twisted pair copper lines corresponding to a number of users are contained in a binder, which ultimately connect the distribution point unit (DPU) to the CPE.  An increase in the frequency increases electromagnetic coupling  which causes significant crosstalk among these closely packed  lines at higher frequencies.   Depending on the position of disturbers with respect to the victim receiver, this crosstalk is classified as FEXT or NEXT, as shown in Fig.\ref{fig:crosstalk_duplexing}. NEXT refers to coupled signals that originate from the same end as the affected receiver. Hence, NEXT is interference between upstream signals and downstream signals from different pairs. As NEXT is not attenuated over the line length, because the interfering  transmitters are close to the receivers, its effect on the receivers is significant. Duplexing techniques such as FDD and TDD are used to avoid the NEXT interference in VDSL and G.fast systems. While, full-duplexing in the single-pair G.(mg)fast systems can double the capacity, it is challenging to work with the self-NEXT limited.    On the other hand, FEXT refers to the coupled signals that originate from the end opposite to that of the affected receiver. Although the FEXT interference decreases with line lengths, its impact remains significant and has adverse effect on the data rate performance.

Another major cause of rate degradation is crosstalk that originates from subscribers enjoying other services, and is referred to as alien crosstalk. Such crosstalk exists in practical situations mainly due to the coexistence  of different DSL services such as G.fast and VDSL in the same binder.  This alien crosstalk manifests itself as spatially correlated (i.e., correlated across lines) additive noise at each tone at the receiver of a vectored system.  Hence, alien crosstalk is  amenable in the vectoring framework by exploiting  spatial correlation between different noise samples of the copper lines.

In downstream transmissions, a CPE can experience interference from other existing broadband services and through the usage of various electrical appliances in the vicinity. The common sources of such interferences are electric power sources, radio frequency interference (RFI)
ingress, interference from home local-area network (LAN) and broadband over powerline communications (PLC).  These interferences can be classified based on the duration as continuous and impulsive, and on occurrence interval such as repetitive and non-repetitive noise sources. The presence
of these external interferences at the CPE causes significant performance degradation and algorithms for its mitigation are desirable.

\begin{figure*}[t]
	\begin{center}
		\includegraphics[scale=0.40]{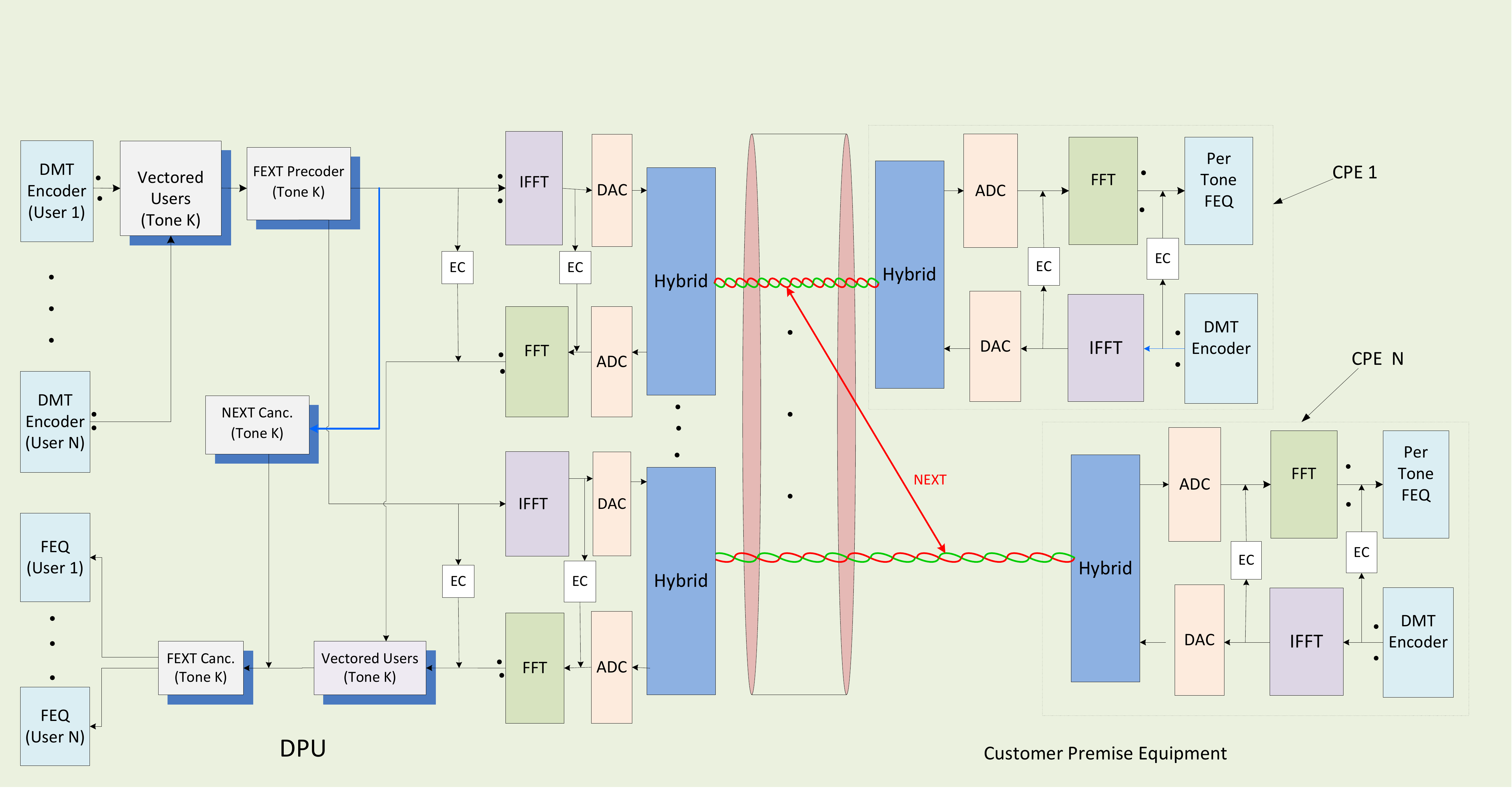}
	\end{center}
	\caption{A schematic diagram of a multi-user multi-carrier DSL system with interference cancellation. There is no signal coordination among CPEs for joint  processing. In the upstream, the received signals from CPEs at each tone are collected at the DPU as a single vector, on which a FEXT canceler is applied. In the downstream, transmit signals for each user at each tone are collected as a single vector at the DPU, and a precoder and NEXT canceler is applied before transmission. At the the CPE, NEXT cancellation is  hard when multiple receiver are co-located without signal-level coordination. Echo canceler (EC) is applied in both time and frequency domains. DMT: discrete multi-tone; FEQ: frequency-domain equalizer; NEXT: near-end crosstalk; FEXT: far-end crosstalk; EC: echo canceler; IFFT: inverse fast Fourier transform; DAC: digital-to-analog converter; ADC: analog-to-digital converter. }
	\label{modem3}
\end{figure*}
\begin{figure*}[t]
	\begin{center}
		\includegraphics[scale=0.55]{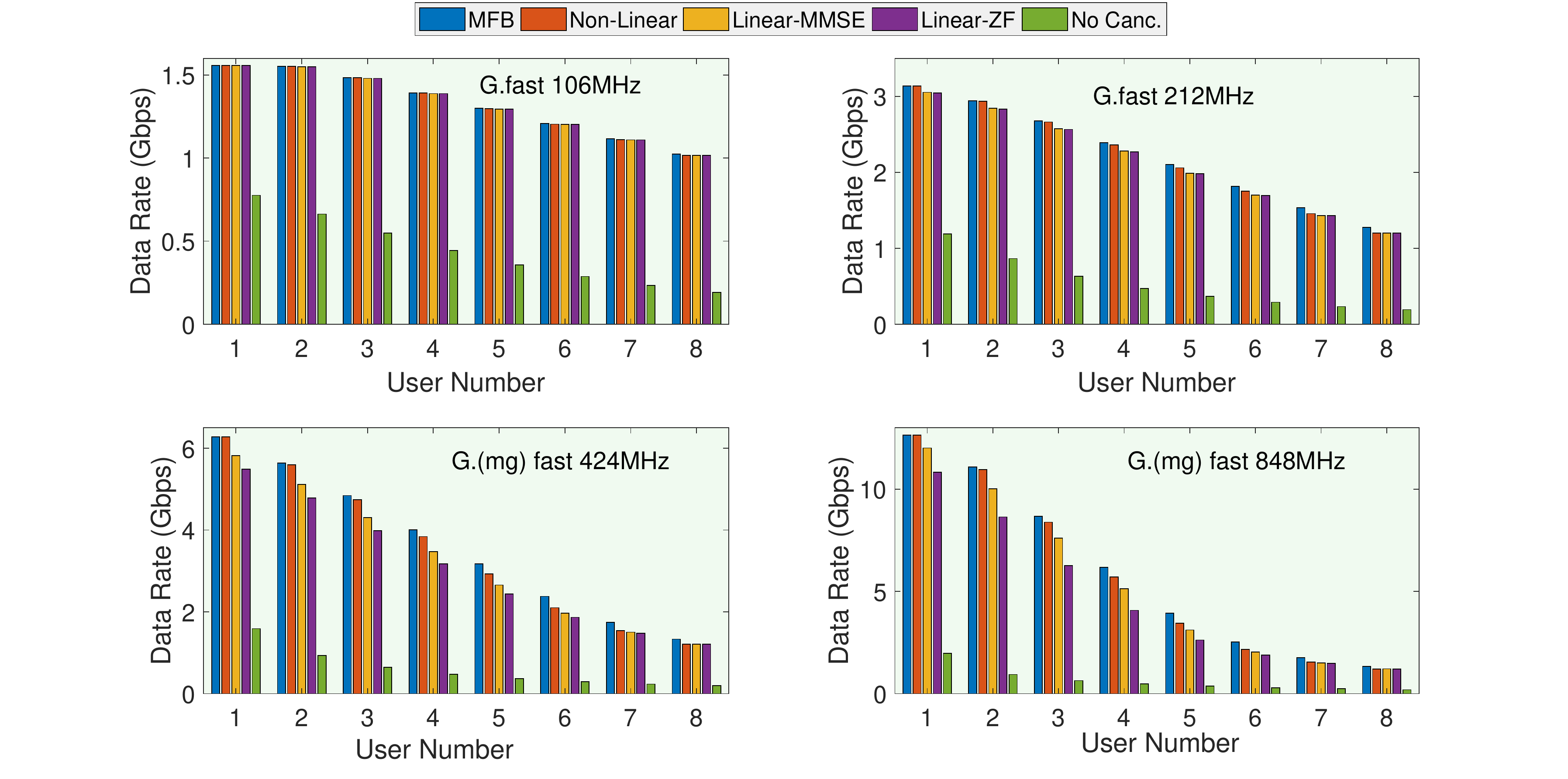}
	\end{center}
	\caption{Simulated achievable data rates of $8$ users in a binder with uniformly spaced line lengths. The first user is  $25$m away from the DPU and the eighth longest line is $200$m. MFB (matched filter bound) is capacity achieved when a single user utilizes both direct	and all FEXT coupling for reception. No Canc. is the capacity achieved  before applying the crosstalk canceler.  The $8\times 8$ vectored channel model is  developed using the direct path of CAT5 cable \cite{g9701} and random crosstalk path as discussed in \cite{zafar2017spm} for the G.fast system. The G.fast channel model is also used for G.(mg)fast performance.  A bit cap of $15$ bits is considered. An SNR gap of $10.75$ \mbox{dB} is taken. The noise power spectral density  is $–140$ \mbox{dBm/Hz}. The transmit signal power spectral density is  $–65$ \mbox{dBm/Hz} for frequencies less than $30$ \mbox{MHz},  $–76$ \mbox{dBm/Hz} between frequencies $30$ \mbox{MHz} and  $106$ \mbox{MHz}, and $–79$ \mbox{dBm/Hz} thereafter. Maximum transmit power constraint of $4$dBm is considered. }
	\label{fig_bar}
\end{figure*}

\begin{figure*}[t]
	\begin{center}
		{\includegraphics[width=\columnwidth]{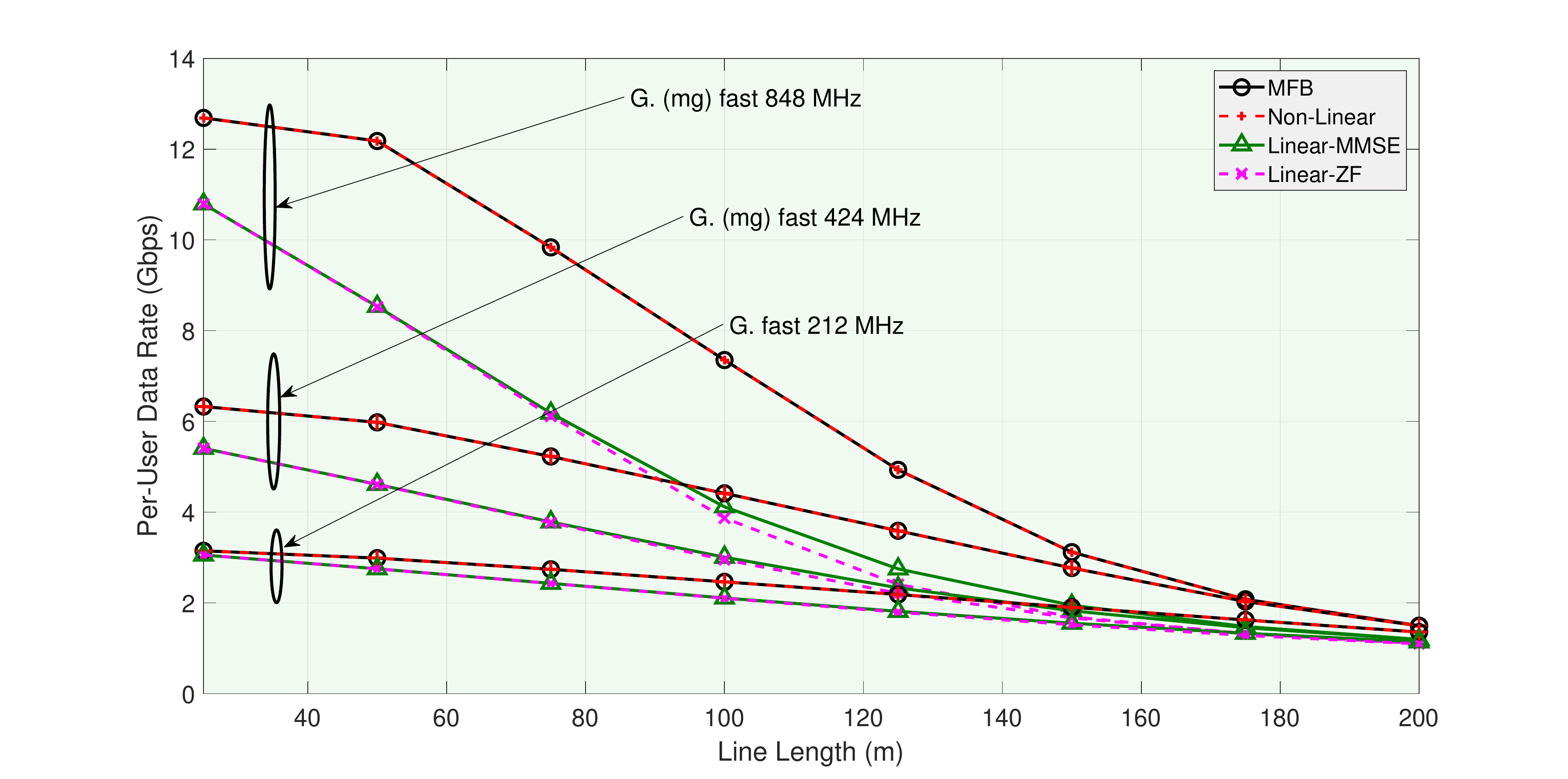}}
		\caption{Average achievable user rate for various crosstalk cancelers. The binder
			is composed of $25$ users with equal line lengths ($20$ \mbox{m} to $200$ \mbox{m}). Other simulation parameters  are described in the caption of Fig.~\ref{fig_bar}.}
		\label{fig:sim_rate25}
	\end{center}
\end{figure*}

\begin{figure*}[t]
	\begin{center}
		\subfigure{\includegraphics[width=\columnwidth]{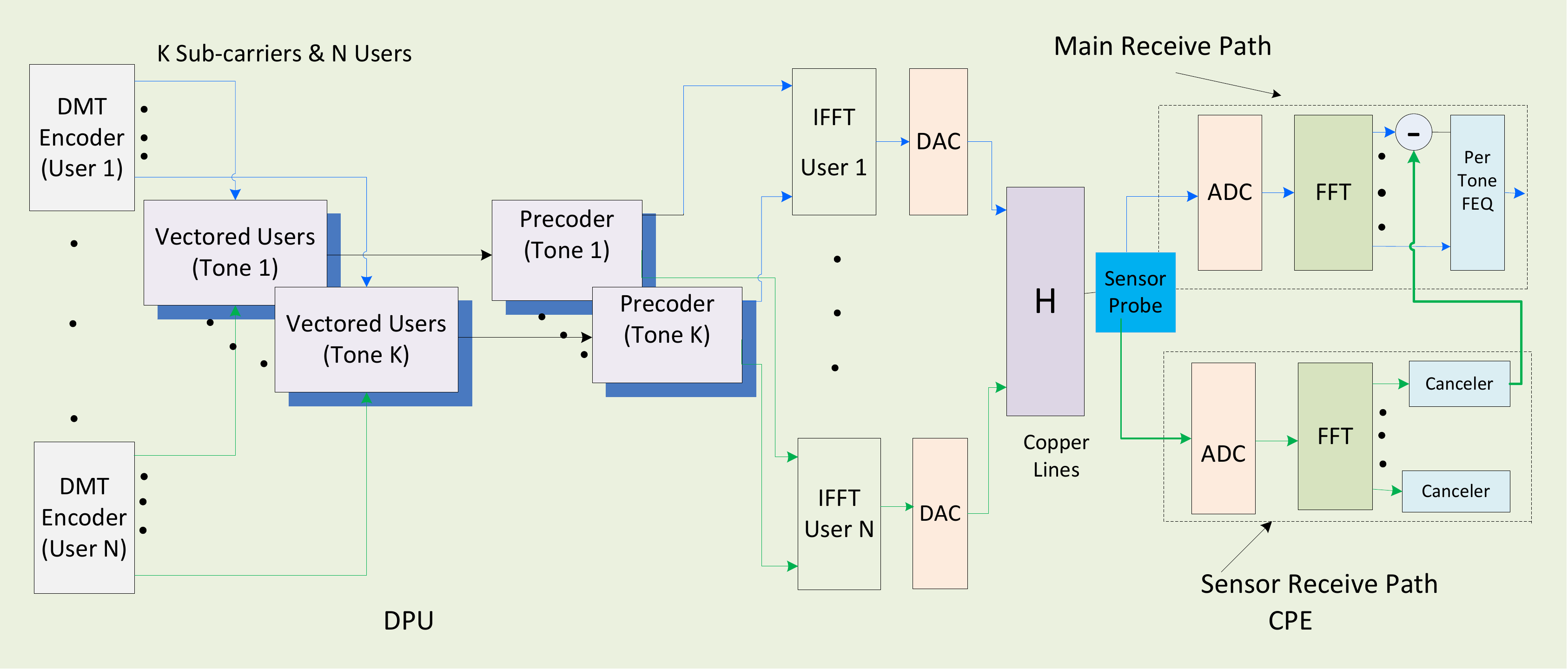}}
		\subfigure{\includegraphics[width=\columnwidth]{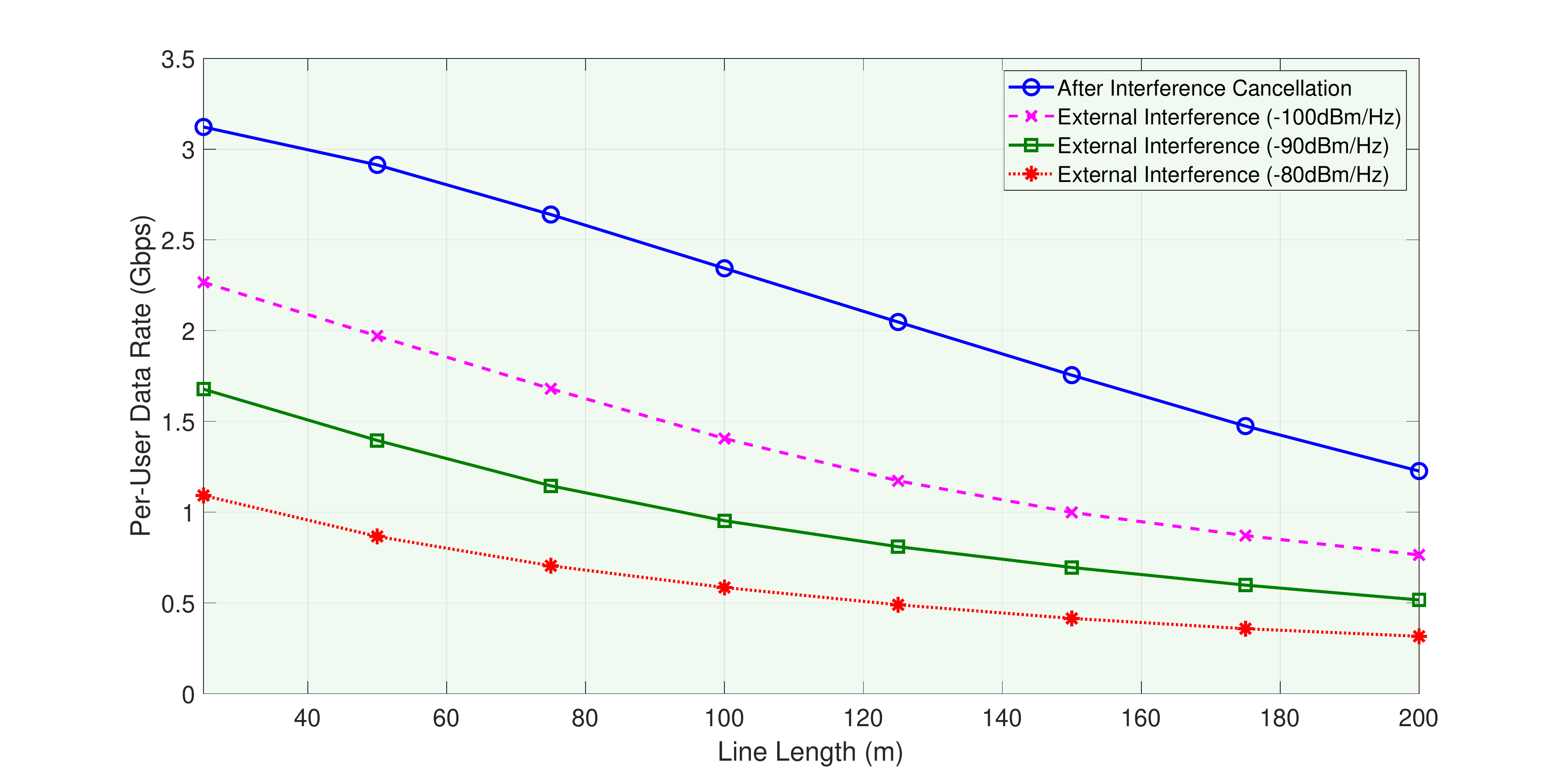}}
		\caption{A sensor based cancellation schematic diagram [left] for external interference at the CPE with vectored precoding for FEXT cancellation in the downstream transmissions for the G.fast system. Transmit signals for each user at each tone are collected as a single vector at the DPU, and a precoder is applied before transmission for pre-compensation of crosstalk. At the CPE, a sensor probe is implemented as a noise reference. The external interference is captured at the sensor which gets  couple to the main receive through a transfer function. The canceler has the task to estimate the coupling transfer function and then use the reference signal to mitigate interference from the main signal. The performance is evaluated [right] for different powers of external interference for the G.fast system $212$ \mbox{MHz} system. The coupling transfer function is frequency dependent and  has an imbalance of $-40$ \mbox{dB}  for frequencies less than $30$ \mbox{MHz},  $–30$ \mbox{dBm/Hz} between frequencies $30$ \mbox{MHz} and  $106$ \mbox{MHz}, and $–20$ \mbox{dBm/Hz} thereafter. }
		\label{modem_cmdm}
	\end{center}
\end{figure*}
\section{Interference Cancellation Techniques}
The capacity of communication over telephone lines is limited by crosstalk. Indeed up to the VDSL2 standard, data rates were below $100$ \mbox{Mbps}. To increase data rates further, joint processing of all lines was used, in a point to multi-point topology.
In contrast to wireless systems where the channels vary very rapidly, the DSL channels vary much slower. This  assists in reducing the amount of training required, and allows interference cancellation of up to $512$ lines in one large system, where both upstream and downstream joint processing are performed at the DPU. The DPU constructs a vector containing all the received/transmit symbols over all lines, and then processes them together to reduce the effect of crosstalk, as shown in Fig.~\ref{modem3}. This vectored processing
is applied at each tone (the DSL systems use $4096$ tones), which is a very intensive computational task. Furthermore, for each vector symbol this processing needs to be done. The design of crosstalk cancelers should achieve the optimal data rate performance while limiting implementation complexity and energy consumption. The main components of the complexity are: computation and tracking of canceler coefficients, the memory requirement to store the  canceler matrices and the application of canceler to the  signal vector.

There several techniques for reducing interference. The simplest are linear techniques, where the transmitter (in the downstream) and receiver (in the upstream) performs a linear combination of the received signals. The simplest approximation was used by \cite{leshem2007} and was used in early VDSL deployment. Next in complexity is the zero forcing (ZF) solution which cancels all interference, by inverting the channel matrix and scaling the diagonal to meet the power constraints.  Next in complexity is the the minimum mean square error (MMSE) linear canceler.  It weights the interference from other lines and the other noises on the channel. These solutions are used upto  G.fast $212$ \mbox{MHz} \cite{Lanneer2017}. Finally, a non linear precoder based on Tomlinson-Harashima precoder (THP) in the downstream or generalized decision feedback equalizer (DFE) in the upstream can be used to prevent power loss in the downstream and noise amplification in the upstream, respectively. This solution is complicated  and has significantly higher complexity but is necessary for G.(mg)fast systems. 
Fig.~\ref{fig_bar} and Fig.~\ref{fig:sim_rate25} depict the data rates achievable using the various interference cancellation/ pre-coding techniques for different technologies.

Finally we would like to mention that in recent years there has been tremendous advances in massive MIMO wireless systems. These solution might find their way into future DSL systems.  Further, we would like to mention that the fact that FEXT is higher than the direct signal suggests that treating the binder as a multiuser MIMO broadcast system in the downstream can lead to better points in the capacity region, where interference is used to amplify the desired signal. This is an interesting area of research for G.(mg)fast since it requires real-time solution of large convex optimization problems. Furthermore, the capacity region in this case is only know under the approximation of Gaussian signaling.

More research is required for efficient crosstalk cancellation on DSL channels at higher frequencies including channel characterization. Iterative processing is another alternative to linear processing with reduced complexity and memory requirements \cite{zafar2018}.  Dynamic spectrum management improves the DSL performance by optimizing power  allocation over the system bandwidth.  
Adaptation and tracking of crosstalk cancelers are important issues and require considerable attention in the context of DSL systems at higher frequencies.

For the mitigation of external interference, use of a noise reference sensor module can be effective at the CPE. Considering a  linear model for external interference, a proof of concept  was developed for VDSL system \cite{zafar_icc2016}.  Data rate performance improvement is significantly higher, as shown in Fig.~\ref{modem_cmdm}. Since the coupling  for the sensor to the main channel is hard to model, deep learning based methods can be used for impulse noise cancellation in DSL systems.

\section{Conclusions}
In this article, we provided an overview of multi-gigabit rate transmission technologies over copper lines with a focus on interference cancellation. We highlighted the salient features of the upcoming DSL technologies and highlighted the key differences from its predecessors. We  discussed various impairments that can be bottleneck in realizing high data rate transmissions. The crosstalk cancellation for  DSL  systems at extremely higher frequency poses new, challenging problems, and further research is needed to fully realize the goal of achieving multi-gigabit data rates over copper lines. Channel modeling for both NEXT and FEXT, and characterization of impulse noise at the higher frequency requires extensive  measurement campaigns and statistical studies. We analyzed the impact of different interference cancellation techniques on the performance of DSL systems, and provided insights on new research direction in this area. A challenging problem  is to deal with interference cancellation in a multi-pair full duplex DSL systems.   Solving this problem, might be feasible only for certain topology and require proper power control. While at the DPU NEXT cancellation is a straightforward implementation of the multichannel echo canceler as is used in Gigabit Ethernet, the problem of the CPE receiver is significant, and data rates will be significantly degraded, especially when multiple receiver are co-located at the same floor.

\bibliographystyle{IEEEtran}
\bibliography{comm_bibtex_file}

\end{document}